\documentclass[twocolumn]{aastex62}
\newcommand{\nf}{\ensuremath{\mathrm{NF}}}
\newcommand{\psra}{\ensuremath{{\rm J}0323+6742}}
\newcommand{\psrb}{\ensuremath{{\rm J}0054+6946}}
\newcommand{\psrc}{\ensuremath{{\rm J}0137+6349}}

\newcommand{\eqref}[1]{(\ref{#1})}

\shorttitle{Gaussian Mixtures for Nulling Pulsars} 
\shortauthors{Kaplan et al.}
 
\submitjournal{ApJ}
\accepted{January 27, 2018}

\begin{document}

\title{A Gaussian Mixture Model for Nulling Pulsars}
\author[0000-0001-6295-2881]{D.~L.~Kaplan}
\affiliation{Center for Gravitation, Cosmology and Astrophysics, Department of Physics, University of Wisconsin--Milwaukee, P.O. Box 413, Milwaukee, WI 53201, USA}
\author[0000-0002-1075-3837]{J.~K.~Swiggum}
\affiliation{Center for Gravitation, Cosmology and Astrophysics, Department of Physics, University of Wisconsin--Milwaukee, P.O. Box 413, Milwaukee, WI 53201, USA}
\author{T.~D.~J.~Fichtenbauer}
\affiliation{Center for Gravitation, Cosmology and Astrophysics, Department of Physics, University of Wisconsin--Milwaukee, P.O. Box 413, Milwaukee, WI 53201, USA}
\author[0000-0002-4162-0033]{M.~Vallisneri}
\affiliation{Jet Propulsion Laboratory, California Institute of Technology,
Pasadena, CA 91109, USA}
\affiliation{TAPIR, California Institute of Technology, Pasadena, CA 91125, USA}

\correspondingauthor{D.~L.~Kaplan}
\email{kaplan@uwm.edu}

\begin{abstract}
The phenomenon of pulsar nulling -- where pulsars occasionally turn off for one or more pulses -- provides insight into pulsar-emission mechanisms and the processes by which pulsars turn off when they cross the ``death line.''  However, while ever more pulsars are found that exhibit nulling behavior, the statistical techniques used to measure nulling are biased, with limited utility and precision.  In this paper we introduce an improved algorithm, based on Gaussian mixture models, for measuring pulsar nulling behavior.  We demonstrate this algorithm on a number of pulsars observed as part of a larger sample of nulling pulsars, and show that it performs considerably better than existing techniques, yielding better precision and no bias.  We further validate our algorithm on simulated data.  Our algorithm is widely applicable to a large number of pulsars even if they do not show obvious nulls.  Moreover, it can be used to derive nulling probabilities of nulling for individual pulses, which can be used for in-depth studies.
\end{abstract}


\section{Introduction}
Pulsar surveys to date have found {$\sim$2,700} pulsars \citep{2016yCat....102034M}\footnote{http://www.atnf.csiro.au/people/pulsar/psrcat}, most of which move across the $P\mbox{--}\dot{P}$ diagram and turn off in $\sim 10^7\,$yr when they pass the ``death line" \citep[see, e.g.,][]{2012hpa..book.....L}.
Just how and why pulsars turn off is still a subject of intense scrutiny, 
with ongoing observational and theoretical investigations 
(see references below).  Nulling pulsars \citep{1970Natur.228...42B}  -- pulsars whose radio 
emission ceases temporarily for one or more rotations -- offer an invaluable 
laboratory to study pulsar emission mechanisms and magnetospheres.
The meaning of null durations, the intervals
between them, and the underlying mechanism
have been matters of debate ever since the behavior was recognized.
  
The first comprehensive study of nulling pulsars \citep{1976MNRAS.176..249R}
suggested that as a pulsar ages, the time interval between
regular bursts of pulsed emission increases, eventually leading to
``death" when the interval between bursts is much longer
than the duration of the bursts themselves. {A later study of 72 pulsars found a stronger
correlation between null fraction
(\nf; the fraction of time that a pulsar spends in a null state) and spin period, 
but still argued that nulling could be indicative of
a faltering emission mechanism \citep{1992ApJ...394..574B}.  \citet{2007MNRAS.377.1383W} argued on the basis of a smaller sample of 45 nulling pulsars that  nulling behavior was more related to a large characteristic age than other parameters, but the analysis was not quantitative.}



However, while data-collection capabilities, processing techniques,
and theories explaining pulsar nulling \citep[e.g.,][]{2012MNRAS.424.1197G} have become more sophisticated,
we are still in a regime where increasing the sample size of nullers
along with the precision of the nulling analysis can have a
significant impact on our understanding.  We are working to increase
the sample size through detailed followup observations of sources
found in the Green Bank North Celestial Cap survey (GBNCC;
\citealt{2014ApJ...791...67S,lynch18,kawash18}).  At the same time, the
classification of pulsars into those that do or do not null is largely
based on qualitative examination by eye, and computation of the nulling
fraction suffers from significant limitations that may bias the
results.

Here we present a robust technique for determining nulling
fractions using Gaussian mixture models
\citep[see, e.g.,][]{2014sdmm.book.....I}.  As we demonstrate, the
robustness of this technique allows us to infer quantitative limits on nulling
fractions even for pulsars that have no obvious nulls, so the technique can be
used for more sophisticated population analyses.  First, we describe
the sample data that we used to test our technique
(\S~\ref{sec:data}).  We then discuss the technique itself
(\S~\ref{sec:gmm}), along with fit results to actual data
(\S~\ref{sec:results}) and simulated data (\S~\ref{sec:simulation}).
We conclude in \S~\ref{sec:conc}.  All of our source code is available
at \dataset[10.5281/zenodo.1155855]{https://doi.org/10.5281/zenodo.1155855}.  Note that we believe a similar algorithm was applied
to nulling pulsars by \citet{2014ASInC..13...79A}, but we have been
unable to find details of its implementation or results.

\section{Sample Data}
\label{sec:data}
To test our method, we gathered timing data from a recent study of new pulsar discoveries in the GBNCC survey \citep{lynch18}. Data were
collected with the 100-m Robert C.\ Byrd Green Bank Telescope (GBT), observing at a center 
frequency of 800\,MHz.  We digitized a  200\,MHz bandwidth using the Green Bank Ultimate Pulsar 
Processing Instrument (GUPPI; \citealt{2008SPIE.7019E..1DD}) in incoherent search mode, 
with 2,048 frequency channels and sampling every 40.96\,$\mu$s. Pulsars included in this study 
(PSRs \psra, \psrb, and \psrc) were observed monthly for 3-minute exposures between 
February 2013 and January 2014.
Exposure times for PSR~\psrb\ were lengthened to 6\,minutes starting in November 2013 through May 2014.

For each pulsar, data were folded modulo the pulse period using 
{\tt DSPSR} and RFI was zapped interactively using
the {\tt pazi} routine from {\tt PSRCHIVE}.
{A crucial step in nulling studies is the choice of \emph{on-pulse} and \emph{off-pulse} windows,
defined as fixed phase intervals (within folded profiles) that are assumed to contain most and none,
respectively, of the pulsar emission. Loosely speaking, the intensities observed in the off-pulse window
are attributed to background noise setting the threshold for on-pulse--window intensities to be classified
as nulling or emitting.
We set the width and phase of on-pulse windows manually by inspecting folded profiles, choosing}
off-pulse windows of the same size at pulse phases containing no visible emission, often about 
0.5 rotations from the on-pulse window, but careful not to include any interpulse (if present). 

\begin{figure}
  \includegraphics[width=\columnwidth]{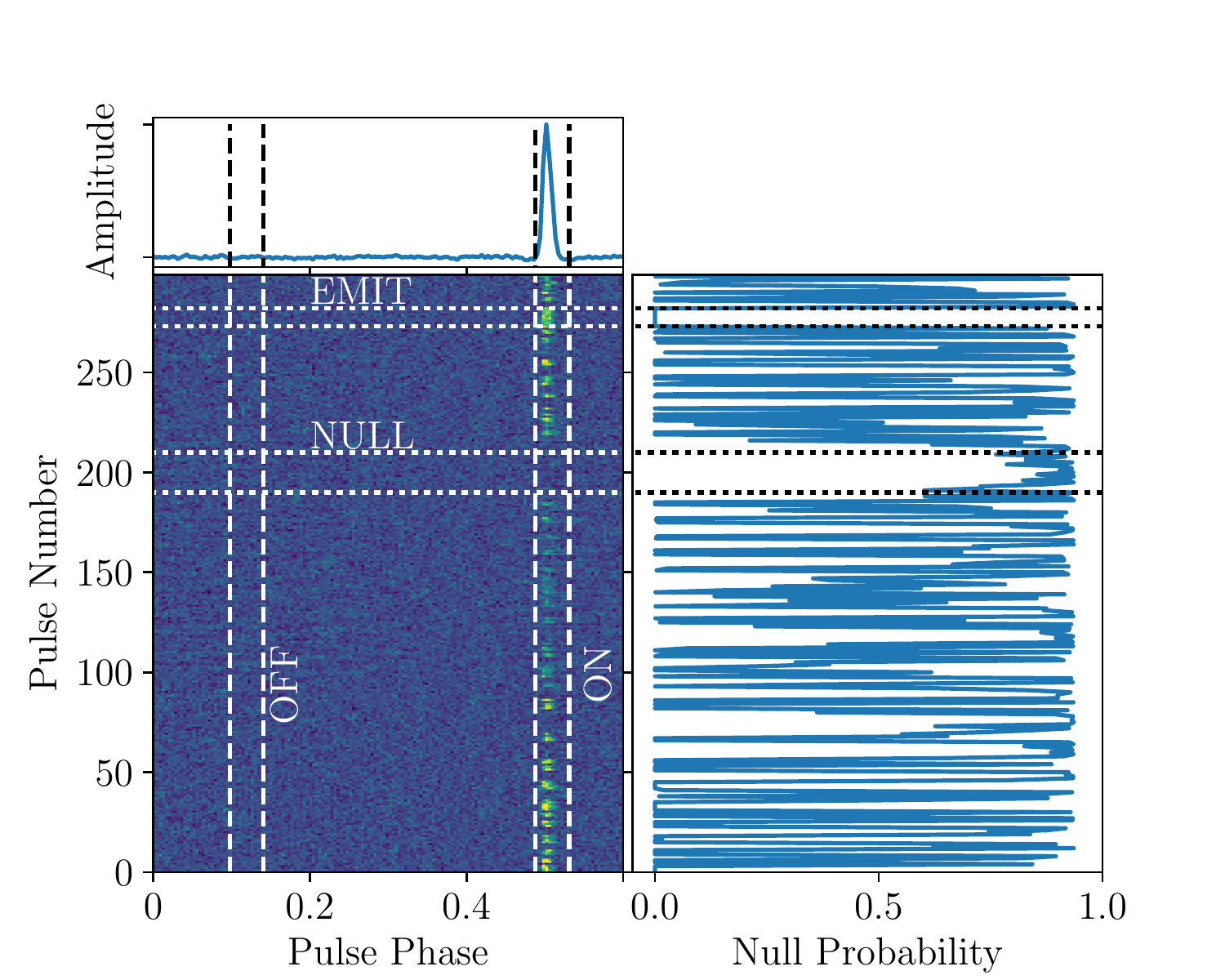}
  \caption{Amplitude versus pulse phase and pulse number for a subset
    of the data on PSR~\psra.  The top panel shows the average
    pulse profile, with the on- and off-pulse phase windows delineated by
    the vertical lines.  The bottom-left panel shows the amplitude for individual pulses, with the phase windows indicated.
    The right panel shows the probability of nulling (Eqn.~\ref{eqn:response}).  {We highlight regions with significant nulling behavior (20 successive pulses with minimum \emph{a posteriori} null probability greater than 62\%) and relatively steady emission (10 pulses with maximum \emph{a posteriori} null probability $\sim$ 0\%) with the horizontal lines.  Such periods of successive nulls or emitted pulses are not intrinsically different from those of pulsars that switch states more rapidly, but are easier to visualize.}
    }
  \label{fig:0325}
\end{figure}

We fitted and removed a 6th-order polynomial to flatten the baseline of each single 
pulse profile {and} center the off-pulse noise on zero intensity (similar to \citealt{Lynch13,Rosen13}).
Afterwards, we assembled our dataset of on and off intensities for each individual pulse 
by integrating over each phase bin and concatenating the results from the  separate 3-min/6-min exposures.
See  Figure~\ref{fig:0325} for an illustration of the pulse intensities measured in on- and off-pulse windows, along with some significant nulling behavior.

\section{A Gaussian Mixture Model for Nulling Pulsars}
\label{sec:gmm}

The standard algorithm  for fitting  the nulling fraction \nf\ \citep{1976MNRAS.176..249R} is to first
construct histograms of the integrated intensities $I^\mathrm{ON} \equiv \{I_k^\mathrm{ON}\}$ and $I^\mathrm{OFF} \equiv \{I_k^\mathrm{OFF}\}$ measured in the on-pulse and off-pulse windows at each observation $k$. We denote the histograms as $\mathrm{ON}_n(I^\mathrm{ON})$ and $\mathrm{OFF}_n(I^\mathrm{OFF})$, where the index $n$ identifies histogram bins.
Then, for a series of trial values of \nf\, one computes the difference histogram
$\Delta_n(I^\mathrm{ON},I^\mathrm{OFF})=\mathrm{ON}_n(I^\mathrm{ON})-\nf\times\mathrm{OFF}_n(I^\mathrm{OFF})$.
The best-fit $\hat \nf$ is the value that minimizes the sum of $\Delta_n(I^\mathrm{ON},I^\mathrm{OFF})$ over negative intensities,
$|\sum_{I_n<0}\Delta_n(I^\mathrm{ON},I^\mathrm{OFF})|$, where the nulling is presumed to dominate.
This method has a number of drawbacks: first, it
requires construction of histograms so there is an arbitrary choice of
binning.  Second, it assumes that the pulses with $I_k^\mathrm{ON}<0$ are entirely
due to nulling, excluding weak pulsars where emitted pulses
be overwhelmed by radiometer noise or otherwise end up at negative intensities.

\subsection{Fitting Algorithm}
\label{sec:algorithm}

In our proposed method, we define the likelihood of the on-pulse dataset $I^\mathrm{ON}$ as a
one-dimensional Gaussian mixture model,
\begin{equation}
\label{eq:likelihood}
p(I^\mathrm{ON}|\{w_j,\mu_j,\sigma_j\}) = \prod_k^N \left( \sum_{j=1}^M w_j {\cal N}(I^\mathrm{ON}_k|\mu_j,\sigma_j) \right),
\end{equation}
where $\mu_j$ and $\sigma_j$ are the means and standard deviations of $M$ normal distributions,
and $w_j$ are their weights in the mixture, which must satisfy $\sum w_j=1$.
Therefore there are $3M-1$ free parameters in this model.  The case with $M=2$ corresponds to the
standard description of nulling with two modes; we order results
so that the $j=1$ component has the lowest mean and describes the nulls,
so $\nf = w_1$.
Note that Eq.\ \eqref{eq:likelihood} implies that the error in the measurement of the intensities
is negligible compared to the intrinsic scatter in the mixture components, described by the $\sigma_j$.\footnote{If the measurement error is not negligible, the $\sigma_j$ are effectively redefined to include it, under the assumptions that the error is similar for every observation.}

The parameters that maximize this likelihood can be found using the 
expectation-maximization algorithm \citep{Dempster77maximumlikelihood},
implemented as \texttt{GaussianMixture} in \texttt{scikit-learn}
\citep{scikit-learn}. Using this implementation, we find reasonable results
for pulsars with significant separations between nulling and emitting pulses.
Such separation may be a signal-to-noise effect, but scatter in pulse intensities could also
result from intrinsic variability in the pulsars themselves or
interstellar scintillation \citep{1990ARA&A..28..561R,2003ApJ...596L.215J}.
However, as we show below the results are not ideal for other pulsars.

We can do better by using the off-pulse dataset $I^\mathrm{OFF}$ to constrain the null-component parameters
$\mu_1$ and $\sigma_1$, by way of the off-pulse likelihood
\begin{equation}
\label{eq:offlikelihood}
p(I^\mathrm{OFF}|\mu_1,\sigma_1) = \prod_k {\cal N}(I^\mathrm{OFF}_k|\mu_1,\sigma_1).
\end{equation}
Indeed, we may think of this step as providing a prior distribution for $\mu_1$ and $\sigma_1$, which is then used in Eqn.~\ref{eq:likelihood}.
We then explore the $\{w_j,\mu_j,\sigma_j\}$ parameter space using Markov Chain Monte Carlo (MCMC) techniques,
specifically the affine-invariant population-MCMC algorithm \texttt{emcee} described by 
\citet{2013PASP..125..306F}.
The details of our implementation are as
follows:
\begin{itemize}
  \item We initialize 40 \texttt{emcee} ``walkers'' around the best-fit region for ${w_j, \mu_j, \sigma_j}$, as
    determined by expectation maximization run on $I_k^{\rm ON}$.
  \item For simplicity, we set the $\mu_1$ and $\sigma_1$ priors as normal distributions centered on $\frac{1}{k} \sum_k I_k^{\rm OFF}$ and on the inner-quartile range of the $I^{\rm OFF}$, respectively, with widths determined following \citet{af03}.
  \item Prior distributions for $\mu_j$ and $\sigma_j$ (with $j>1$) are taken to be flat.  The prior distribution for the weights $w_j$ is a Dirichlet distribution, but since the sum of the weights is 1 it is effectively flat.
  \item We run the walkers through 50 iterations to achieve ``burn in.''
  \item Last, we run the walkers for 500 iterations to obtain the final population, representative of the $\mu_j$, $\sigma_j$, and $w_j$ joint posterior.
\end{itemize}
We experimented with increasing the number of walkers and iterations and found that the values above gave sufficiently reliable results for data-sets with a few thousand pulses and $M<5$, but they can be increased as needed to achieve reliable posterior distributions.

\subsection{Fit Results}
\label{sec:results}

\begin{figure}
   \includegraphics[width=\columnwidth]{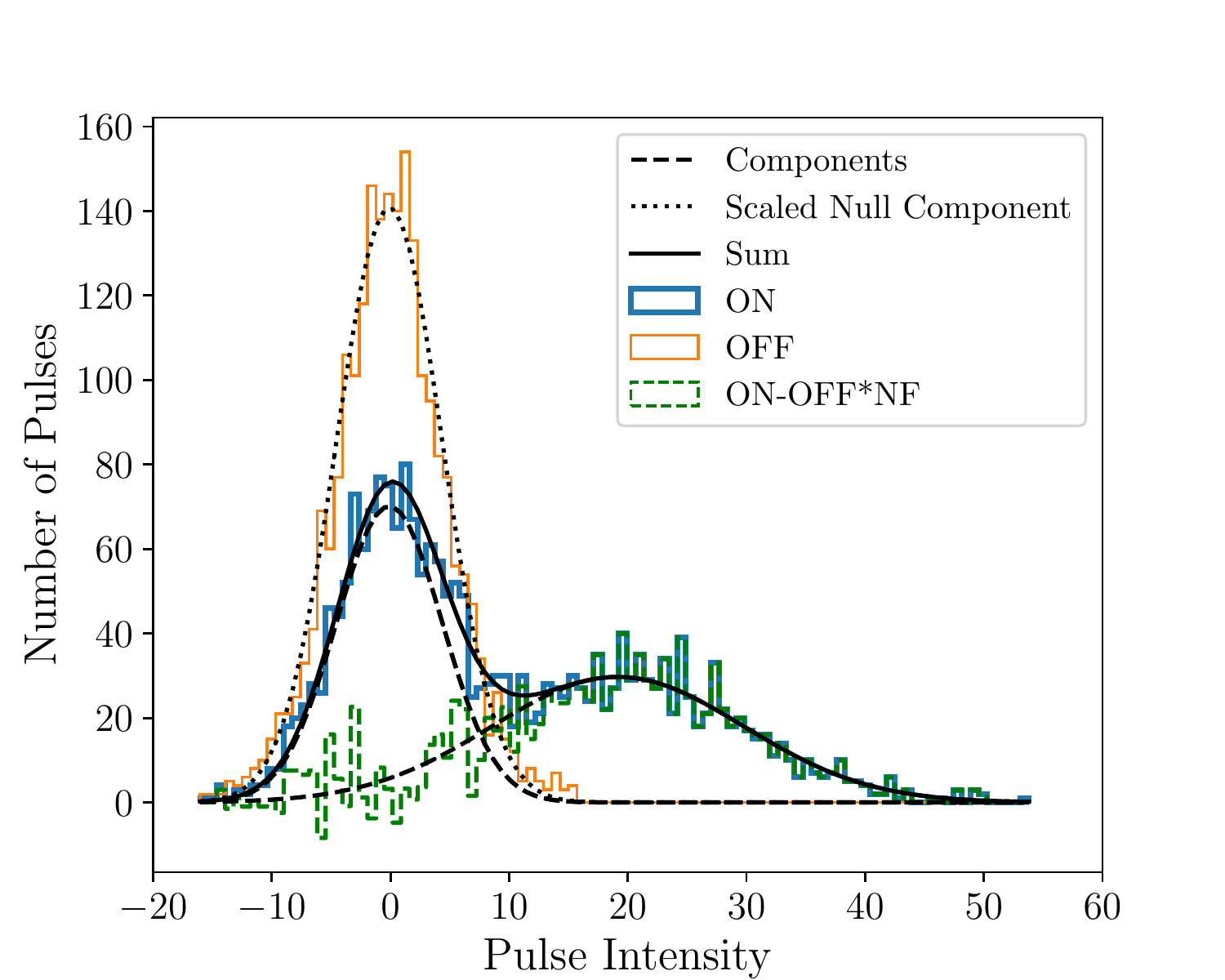}
  \caption{Distribution of pulse intensities for PSR~\psra.  The
    blue and orange histograms are the raw intensities for on- and
    off-pulse windows, respectively.  The dashed curves are the
    maximum \emph{a posteriori}
    individual components from the Gaussian mixture model for $M=2$,
    with the solid curve their sum, as determined by our MCMC algorithm.
    The dotted curve is the component for the nulls scaled by $1/\nf$:
    it matches the off-pulse intensities well.  Finally, the dashed
    green histogram is the data that would be used to implement the
    \citet{1976MNRAS.176..249R} algorithm, although it is plotted for our best-fit value of $\nf$.  Using our MCMC algorithm we find $\hat \nf=50\pm2$\%, compared
  to 56\% using \citet{1976MNRAS.176..249R}.}
  \label{fig:0325hist}
\end{figure}

\begin{figure}
  \includegraphics[width=\columnwidth]{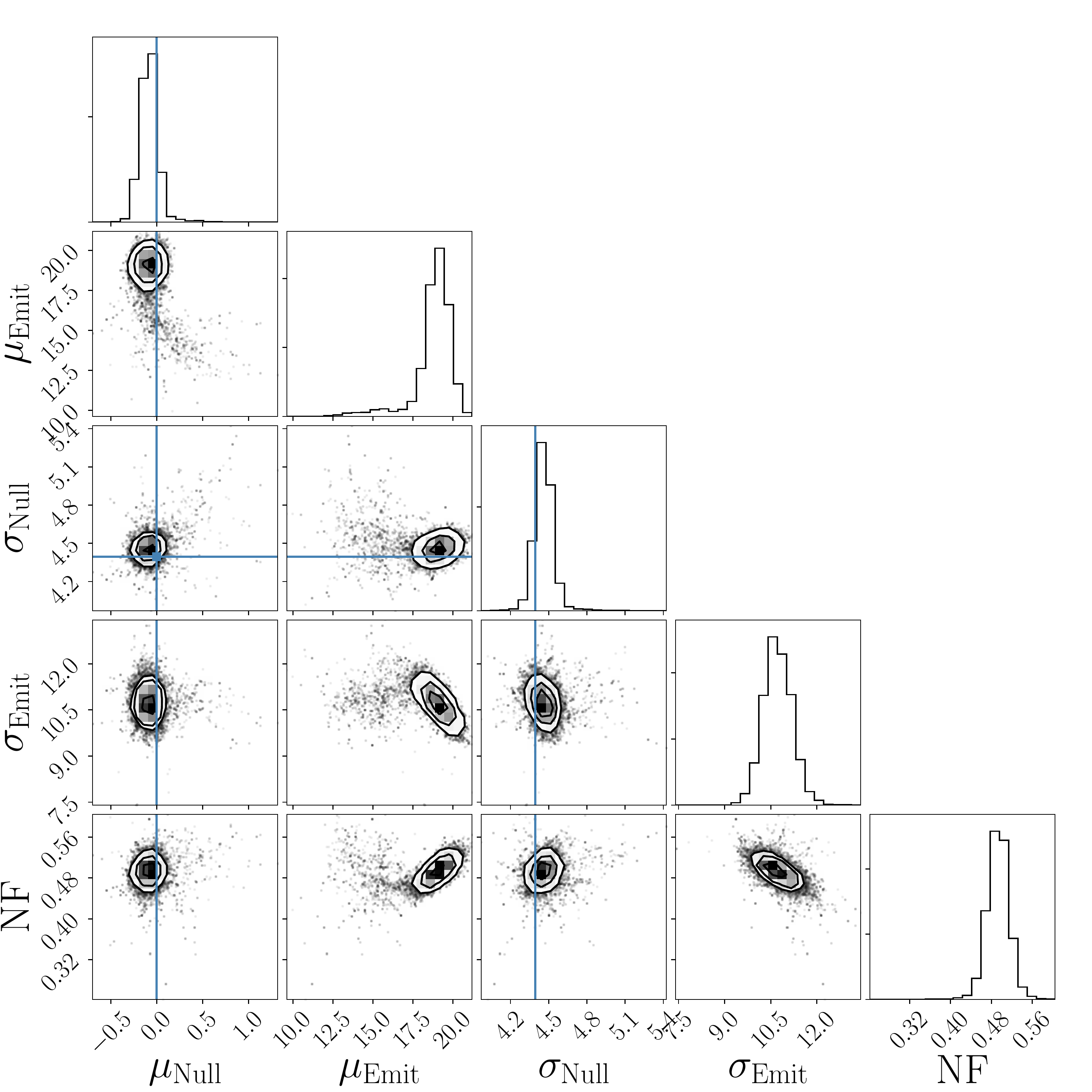}
  \caption{Posterior probability densities for the Gaussian-distribution
  parameters of the null and emitting components, as derived in our MCMC
  algorithm. The solid lines show the $\mu_1$ and $\sigma_1$ modes
  as inferred from the off-pulse data.  The contours are 1-, 2-, and 3-$\sigma$ joint confidence contours.
}
  \label{fig:0325mcmc}
\end{figure}

Representative results from this algorithm are shown in Figure \ref{fig:0325hist},
where we plot $p(I^\mathrm{ON}|\{\hat{w}_j,\hat{\mu}_j,\hat{\sigma}_j\})$ (solid line) and
$p(I^\mathrm{OFF}|\hat{\mu}_1,\hat{\sigma}_1)$ (dashed)
on top of the $I^\mathrm{ON}$ and $I^\mathrm{OFF}$ histograms (with appropriate normalizations);
here $\{\hat{w}_j,\hat{\mu}_j,\hat{\sigma}_j\}$ are the \emph{a posteriori} joint maxima
of the Gaussian mixture parameters. 
Figure \ref{fig:0325mcmc} shows the posterior densities of the Gaussian means and variances.
The data appear to be fit well, with the null component ending up close to the off-pulse fit results; we observe no significant pathologies in the MCMC chains.
{To evaluate goodness of fit quantitatively, we perform the Kolmogorov--Smirnov test (KS test; \citealt{chakravarti1967handbook}), and
find a statistic value of 0.015, corresponding to a $p$-value of 0.7---no evidence to reject the hypothesis that the data were sampled from the best-fit distribution.}

For this pulsar we find that the emitting 
and the null components are sufficiently separate that all of the
algorithms outlined above would give similar results.  For instance,
we find that only about 2\% of the pulses from the emitting
component would have intensities less than 0, which would only bias the
\citet{1976MNRAS.176..249R} results by a small amount.

\begin{figure}
  \includegraphics[width=\columnwidth]{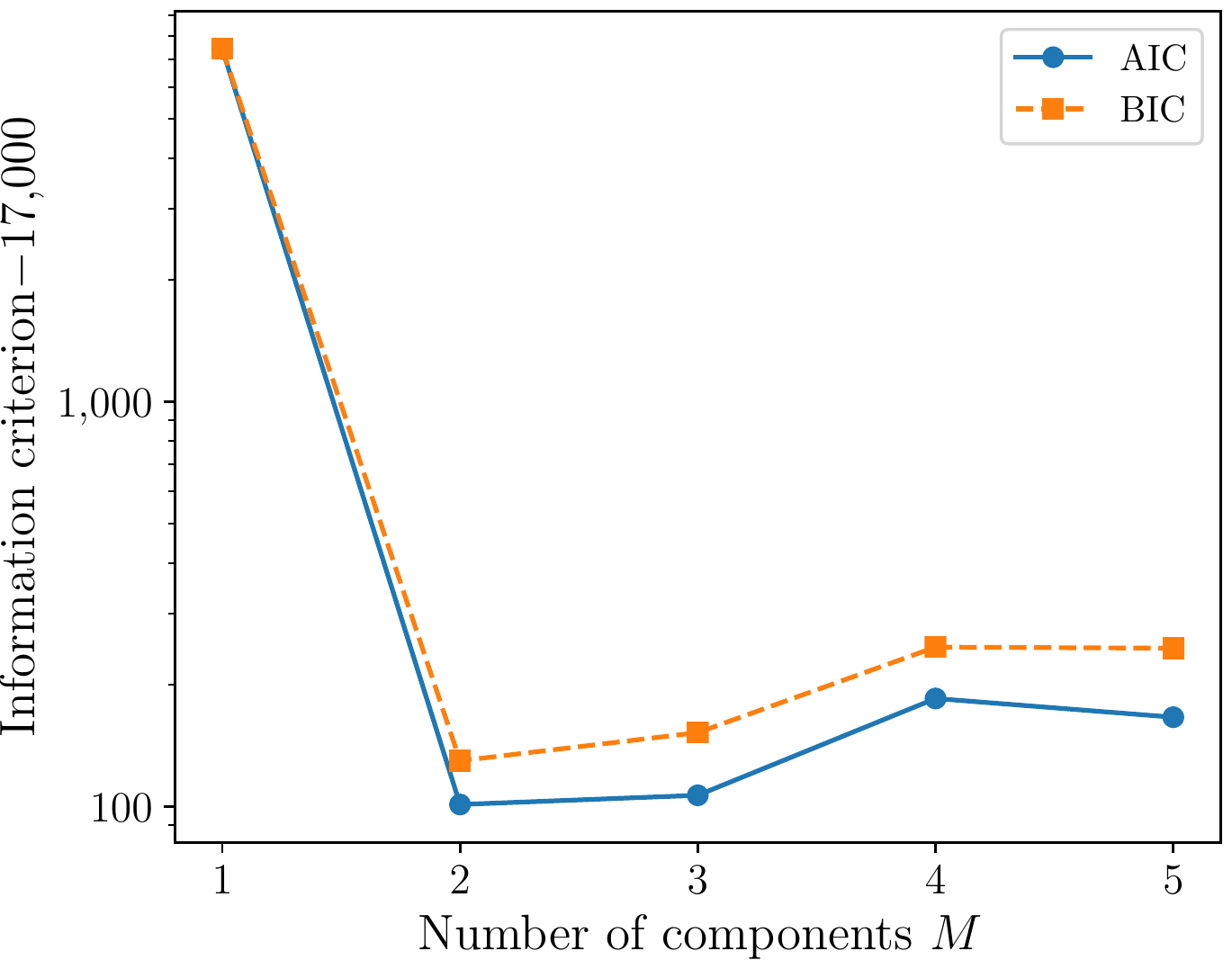}
  \caption{Model-comparison evidence for different number of Gaussian-mixture components
  in the PSR~\psra\ data.  We show the
    Bayesian information criterion (BIC; orange squares) and the Akaike information
    criterion (AIC; blue circles) as a function of the number of components $M$.
    The non-nulling hypothesis ($M=1$) is rejected at high confidence,
    and $M=2$ is preferred by both criteria. }
  \label{fig:0325models}
\end{figure}

If the problem is well behaved,
we can select the optimal number of components $M$ by maximizing the
Bayesian information criterion (BIC; \citealt{schwarz1978}) or the
Akaike information criterion (AIC; \citealt{akaike74}).
We illustrate this in Figure~\ref{fig:0325models}, which shows both.
We find strong evidence that nulling
behavior is present ($M>1$), and a weaker preference for $M=2$ compared to $M>2$.
The BIC corresponds to approximating the Bayes ratios between models as
$e^{-\Delta_{\rm BIC}/2}$, where $\Delta_{\rm BIC}$ is the difference in BIC (likewise for AIC).
In this approximation, the implied Bayes ratio for $M=2$
vs.\ $M=3$ is $O_{23}=175$ (AIC) or $O_{23}=92,000$ (BIC), showing that
$M=2$ is indeed preferred.

\begin{figure}
  \includegraphics[width=\columnwidth]{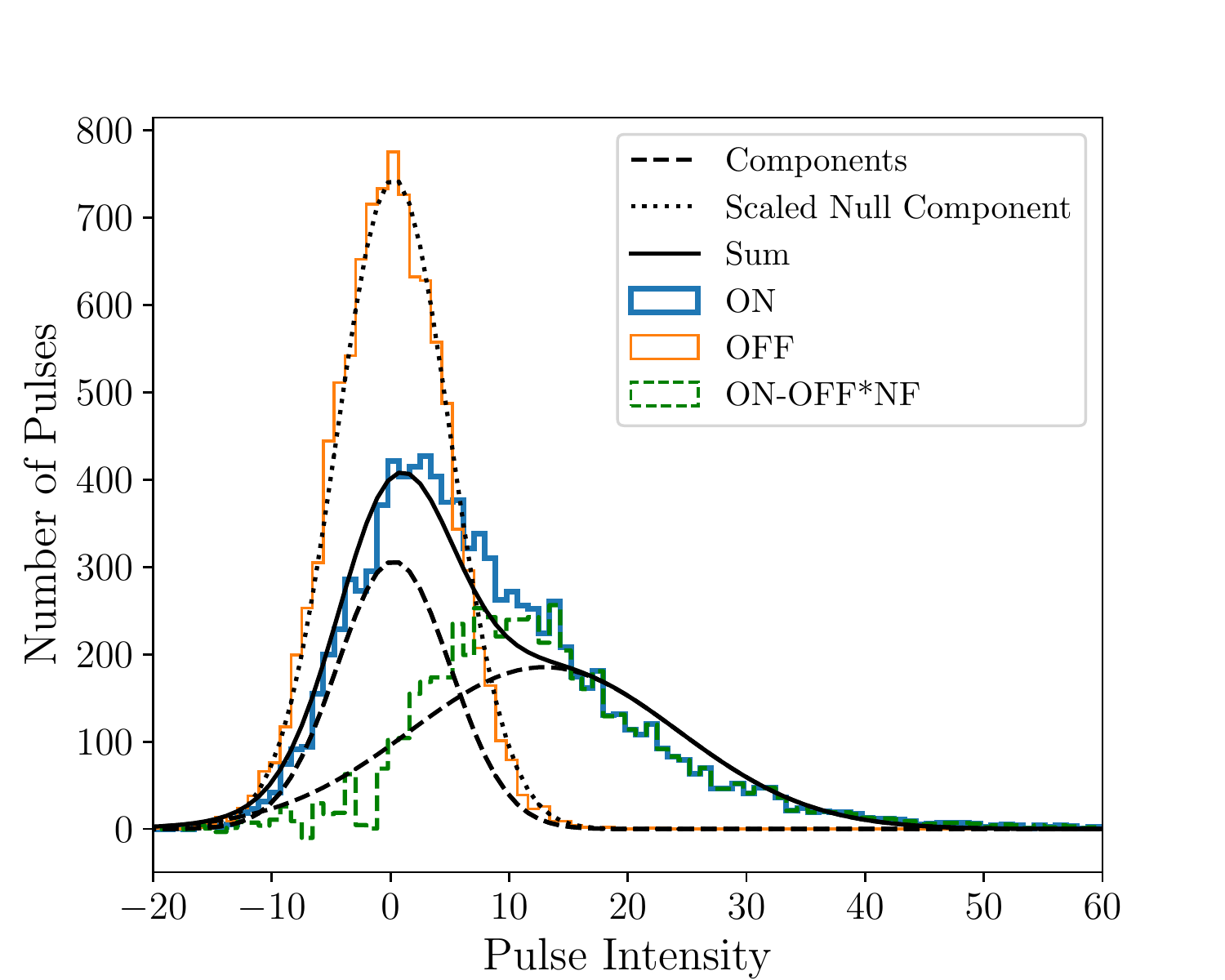}
  \caption{Distribution of pulse intensities for PSR~\psrb\ (see
    Fig.~\ref{fig:0325hist} for details).
    Here the nulling and emitting components are much less separated than in Figure \ref{fig:0325mcmc}.
	Using our MCMC algorithm we find $\hat \nf=40.8\pm1.4$\%, compared
  to 48\% using \citet{1976MNRAS.176..249R}.}
  \label{fig:0053hist}
\end{figure}

We show another example in Figure~\ref{fig:0053hist}, where for $M = 2$ we 
find on-pulse maximum \emph{a posteriori} parameters
$\hat{\mu}_2=12.9$ and $\hat{\sigma}_2=11.3$,
with $\hat{\nf}=41$\%.  In this case there is much less separation in
intensities between nulls and pulses, with about 13\% of the
pulses having $I<0$.  Nonetheless our method gives a robust fit.
Figure~\ref{fig:0053hist} appears to show slight deviations from Gaussian distributions,
which may be handled with more complex models that (for
example) incorporate asymmetric distributions due to scintillation. {Indeed, the KS test rejects the assumption that the data were drawn from the best-fit distribution at the $5\times 10^{-12}$ level; nevertheless, since our goal is primarily to quantify the bimodality of the emitted pulses rather that the exact intensity distribution, we believe the nulling results themselves to be robust.}

\begin{figure}
  \includegraphics[width=\columnwidth]{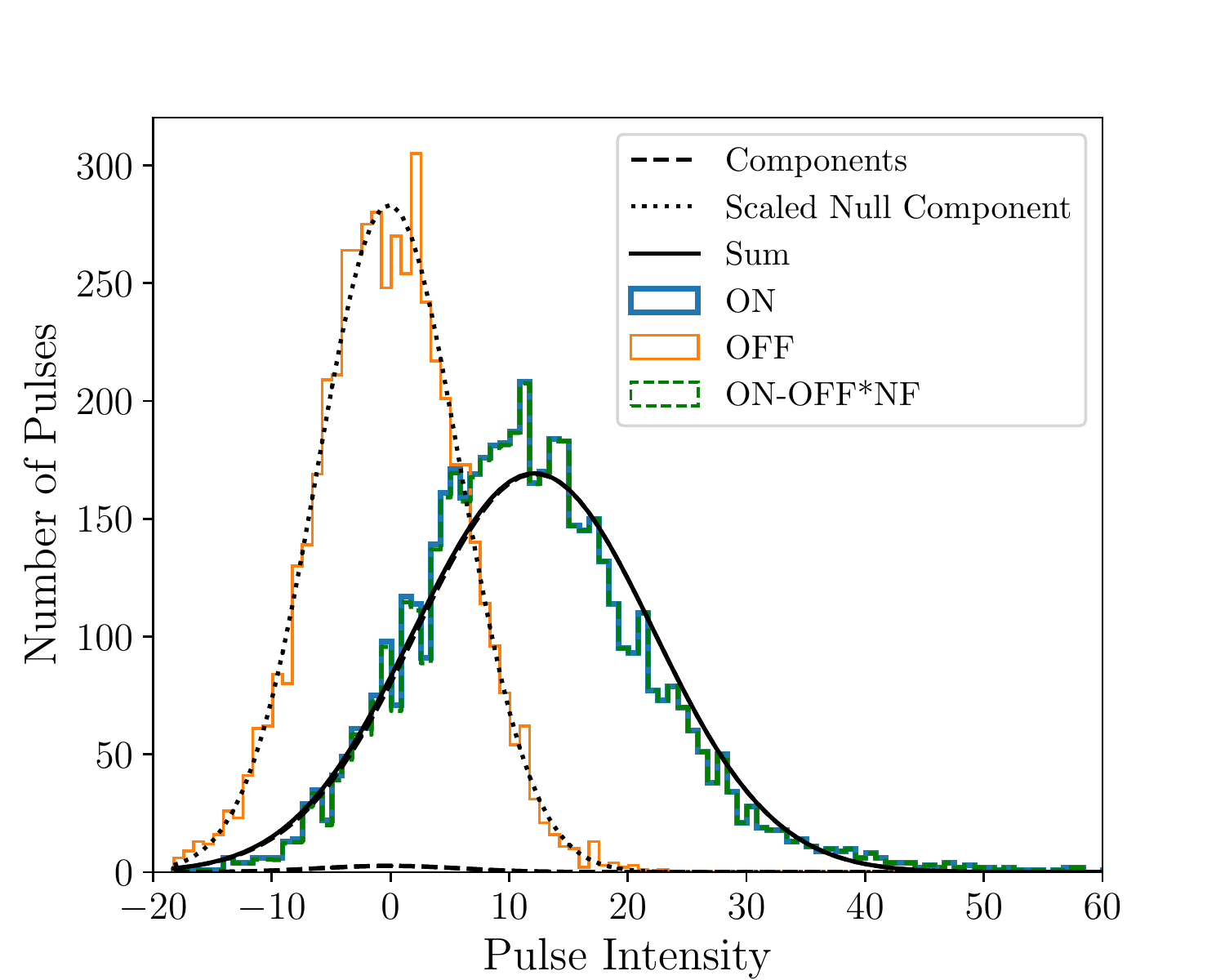}
  \caption{Distribution of pulse intensities for PSR~\psrc\ (see
    Fig.~\ref{fig:0325hist} for details).  Here there is no obvious
    evidence for nulling.  Using our MCMC algorithm we find $\hat \nf=0.9\pm6.4$\%, compared
  to 21\% using \citet{1976MNRAS.176..249R}.}
  \label{fig:0136hist}
\end{figure}

Finally, in Figure~\ref{fig:0136hist}  we show an example where the
pulsar shows no obvious nulling.  Our algorithm finds
$\hat \nf=0.9\pm6.4$\%, consistent with 0, but the
\citet{1976MNRAS.176..249R} algorithm still returns a non-zero value
of 21\%.  Again we see deviations from Gaussian distributions {(with a KS-test $P$ value of $2\times 10^{-7}$)},
but the overall robustness of our determination is evident.  In contrast to
\citet{1976MNRAS.176..249R} which can give  determinations of
non-zero nulling fractions even for pulsars that do not appear to
null, our algorithm behaves well.  Therefore we can use it for all
pulsars that are sufficiently bright regardless of whether nulling is
evident, and derive more robust determinations of whether or not weak
nulling behavior is present.

\subsection{Simulation Results}
\label{sec:simulation}
We validate our algorithm by simulating pulsar data for a range of parameters, drawing random intensities according to Eqs.\ \eqref{eq:offlikelihood} and \eqref{eq:likelihood} for the off- and on-pulse windows, respectively. We base our synthetic datasets on the PSR~\psra\
data analyzed above: we simulate 2,000 pulses with  $\mu_1=0$, $\sigma_1=5$, $\sigma_2=10$, and nulling fraction $\nf=0.5$. We vary $\mu_2$ between 5 (hard to
distinguish from the nulls) and 30 (easily distinguishable), and we repeat the test for 30 trials for each value of $\mu_1$.

\begin{figure}
  \includegraphics[width=\columnwidth]{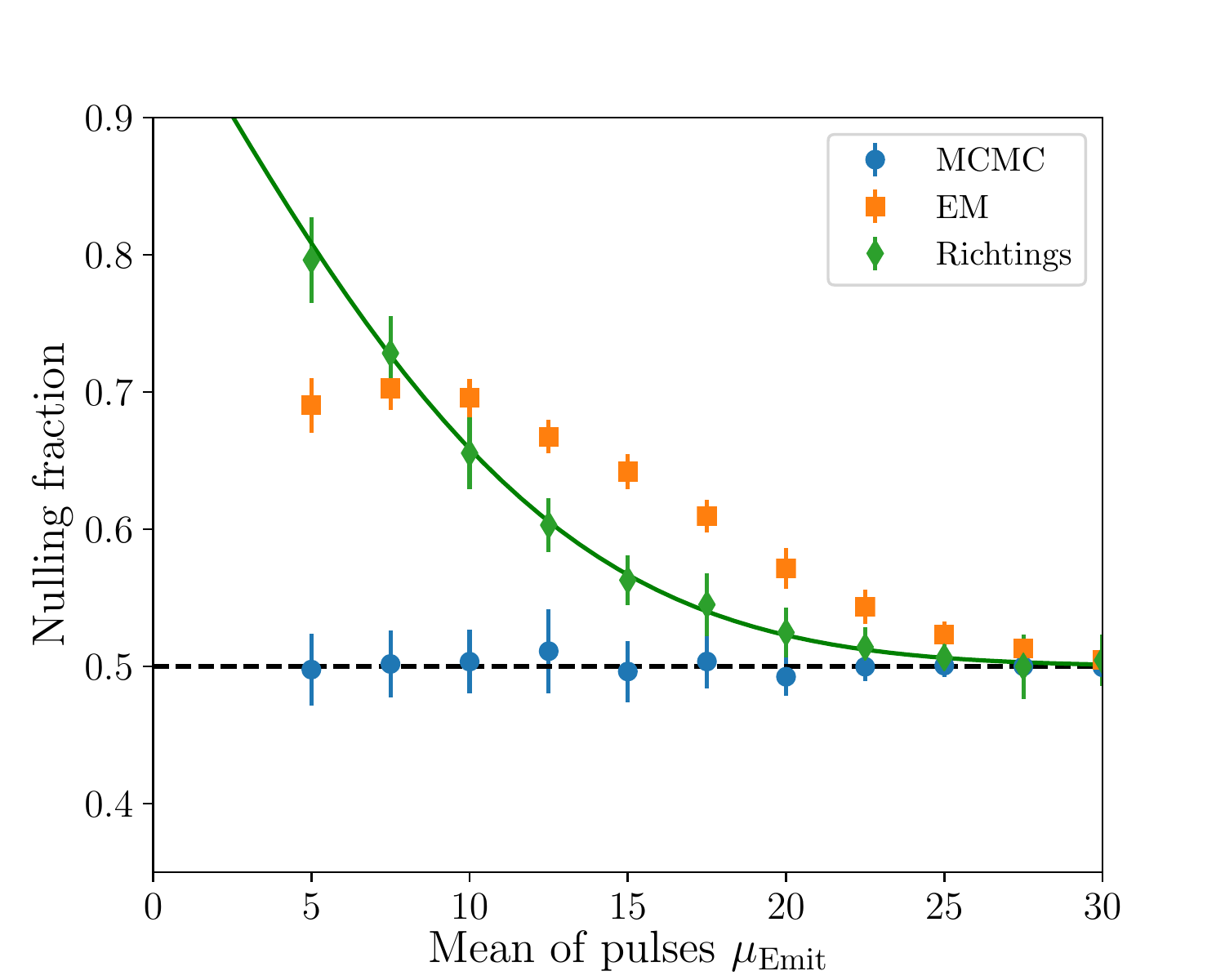}
  \caption{Comparison on \nf\ estimates for simulated data, as derived using the
    \citet{1976MNRAS.176..249R} algorithm (green diamonds), our
    expectation-maximization algorithm (orange
    squares, labeled as ``EM''), and our Bayesian algorithm fit (blue circles).
    The horizontal line marks the true nulling
    fraction of $0.5$. Each simulated dataset consisted of 2,000 pulses, with $\mu_1=0$, $\sigma_1=5$, $\sigma_2=10$; simulations were repeated 30 times for each value of $\mu_2$ (hence the vertical error bars). The solid green curve shows our analytical expectation for the bias of the
  \citet{1976MNRAS.176..249R} algorithm (see main text).}
  \label{fig:sim}
\end{figure}

We plot the median and standard deviation of the estimated \nf\ in Figure~\ref{fig:sim}, using the \citet{1976MNRAS.176..249R} method, the expectation-maximization method, and our Bayesian algorithm (in which case we report the maximum \emph{a posteriori} \nf).
All three algorithms agree for high pulse intensities, $\mu_2 \gtrsim 25$.

We see
that the \citet{1976MNRAS.176..249R} \nf\ estimates are highly biased for low values
of $\mu_2$, as expected.  This is because the model has a
significant fraction of non-nulled pulses with intensities less than
0, ranging from 30\% (for $\mu_2=5$) to 0.1\% (for $\mu_2=30$).  Specifically, we expect a fraction
\begin{equation}
\frac{1}{2}-\frac{1}{2}{\rm erf}\left(\frac{\mu_2\sqrt{2}}{2\sigma_{2}}\right)
\end{equation}
of the emitted pulses to have intensities $<0$, where ${\rm erf}(x)$ is the error function of $x$.  This then leads to a biased estimated \nf,
\begin{equation}
\nf + (1 - \nf)\left[1-{\rm erf}\left(\frac{\mu_2\sqrt{2}}{2\sigma_2}\right)\right]
\end{equation}
which we have plotted in Figure~\ref{fig:sim}, where they agree with
our simulated results.
The EM results using \texttt{GaussianMixture}  are also biased
at low pulse intensities. By contrast, our Bayesian algorithm performs
well, with consistent uncertainties and no obvious bias across the $\mu_2$ range.  


\section{Discussion and Conclusions}
\label{sec:conc}

We have outlined and demonstrated an improved method to determine the nulling fraction of a pulsar. The method performs well in the limit of weak nulling, so it can be applied to a large number of pulsars without evident strong nulls. Unlike the traditional \citet{1976MNRAS.176..249R} algorithm, our method is unbiased, and it can be applied to pulsars with more than two emission modes, as long as those are reflected in the pulse intensities.  However, it does require specification of the functional form of the intensity distributions for the nulling and emitting components: here we assume sums of Gaussians, although exponentials appropriate for 100\% modulation by interstellar scintillation \citep[e.g.,][]{1990ARA&A..28..561R}, or intermediate distributions are also possible.  In those cases the AIC/BIC values can be used to quantitatively compare how well alternative distributions fit the data.

An additional benefit to this analysis is that we can determine explicitly the probability that any individual pulse belongs to a given class.  This is sometimes called the ``responsibility'' \citep{hastie2009elements}, and is given by:
\begin{equation}
p(j | I^{\rm ON}_k) = \frac{w_j {\cal N}(\mu_j, \sigma_j)}{\sum_{j^\prime=1}^M w_{j^\prime} {\cal N}(\mu_{j^\prime}, \sigma_{j^\prime})}
\label{eqn:response}
\end{equation}
for class $j$.
An example of this is shown in Figure~\ref{fig:0325}, where we can determine the nulling probability as $p(j=1 | I^{\rm ON}_k)$.  
This probability can be computed for the maximum \emph{a posteriori} $\{\hat w_j, \hat \mu_j, \hat \sigma_j\}$, or it can be marginalized over their distributions.
Individual-pulse nulling probabilities can be used in robust multi-wavelength studies, to establish whether the X-ray properties of the pulses received during nulls differ from the
others \citep[e.g.,][]{2013Sci...339..436H}.  We can also look for temporal patterns in the nulling
properties, like the length of nulls (Fig.~\ref{fig:0325}) or the time between nulls
\citep[e.g.,][]{2007MNRAS.377.1383W} using quantitative probability
thresholds, and we could examine the  probability that adjacent pulses
transition between nulling and emitting behavior. Finally, we can fit for more than two components and  identify mode changing quantitatively, in addition to nulling. All of these topics will be
explored in future papers. 

\acknowledgements \emph{Acknowledgments.}
We thank S.~McSweeney for helpful comments.  We thank an anonymous referee and the AAS journals statistics editor for their suggestions.
The Green Bank Observatory is a facility of the National Science Foundation operated under cooperative agreement by Associated Universities, Inc. Support was provided by the NANOGrav NSF Physics Frontiers Center award number 1430284. MV acknowledges support from the JPL RTD program. Portions of this research were carried out at the Jet Propulsion Laboratory, California Institute of Technology, under a contract with the National Aeronautics and Space Administration.

\facility{GBT}
\software{Astropy \citep{2013A&A...558A..33A}, DSPSR (\url{http://dspsr.sourceforge.net/index.shtml}), emcee \citep{2013PASP..125..306F}, PSRCHIVE \citep{hotan04},
  scikit-learn \citep{scikit-learn}}.


\end{document}